\newcounter{myctr}

\documentclass{ws-acs}
\usepackage{url}
\usepackage{lscape}

\begin{document}
\makeatletter
\def\@biblabel#1{[#1]}
\makeatother

\markboth{Krzysztof Suchecki, Almila Akdag, Cheng Gao, Andrea Scharnhorst}{Evolution of Wikipedia's Category Structure}

%
\catchline{}{}{}{}{}
%

\title{EVOLUTION OF WIKIPEDIA'S CATEGORY STRUCTURE}

\author{KRZYSZTOF SUCHECKI}
\address{The e-Humanities Group, Royal Netherlands Academy of Arts and Sciences (KNAW)\\
Joan Muyskenweg 25, 1096 CJ Amsterdam, The Netherlands\\
and Erasmus Virtual Knowledge Studio, Erasmus University Rotterdam\\
Burgemeester Oudlaan 50, 3062 PA Rotterdam, The Netherlands\\
currently: IFISC, Instituto de F\'{\i}sica Interdisciplinar y Sistemas Complejos (CSIC-UIB),\\
Campus Universitat Illes Balears, E-07122 Palma de Mallorca, Spain\\
ksucheck@ifisc.uib-csic.es}

\author{ALKIM ALMILA AKDAG SALAH}
\address{UvA, New Media Department, Amsterdam, The Netherlands\\
and Sabanci University, Visual Arts and Communication Design Department, Istanbul, Turkey\\
a.a.akdag@uva.nl}

\author{CHENG GAO}
\address {Room 422, No. 10 Building,  Wencui Rd., Ganjingzi district, Dalian, China}

\author {ANDREA SCHARNHORST}
\address {Data Archiving and Networked Services Department, Royal Netherlands Academy of Arts and Sciences (KNAW), The Hague, The Netherlands\\
and The e-Humanities Group, Royal Netherlands Academy of Arts and Sciences (KNAW)\\
Joan Muyskenweg 25, 1096 CJ Amsterdam, The Netherlands\\
andrea.scharnhorst@dans.knaw.nl}

\maketitle

\begin{history}
\received{(received date)}
\revised{(revised date)}
\end{history}

\begin{abstract}
Wikipedia, as a social phenomenon of collaborative knowledge creating, has been studied extensively from various points of views. The category system of Wikipedia, introduced in 2004, has attracted relatively little attention. In this study, we focus on the documentation of knowledge, and the transformation of this documentation with time. We take Wikipedia as a proxy for knowledge in general and its category system as an aspect of the structure of this knowledge. We investigate the evolution of the category structure of the English Wikipedia from its birth in 2004 to 2008. We treat the category system as if it is a hierarchical Knowledge Organization System, capturing the changes in the distributions of the top categories. We investigate how the clustering of articles, defined by the category system, matches the direct link network between the articles and show how it changes over time. We find the Wikipedia category network mostly stable, but with occasional reorganization. We show that the clustering matches the link structure quite well, except short periods preceding the reorganizations.
\end{abstract}


\section{Introduction}
\label{sect_intro}

One cannot analyze \emph{the Wikipedia} without understanding the technology it is based on. A \emph{wiki} is a software that enables end-users to modify, edit and delete parts of a website via a user-friendly interface. The first of its kind was introduced in 1994, by  Ward Cunningham, who also bestowed the first wiki website its name: WikiWikiWeb. Today wikis are a part of everyday life, especially after the successful implementation of the software by the Encyclopedia Wikipedia initiative in 2000.

Wikipedia has become one of the most used, and most visited free-sources on the Internet, overtaking the expert-based knowledge sources such as Encyclopedia Britannica and the like. This collaborative knowledge creation space has been scrutinized from many perspectives: both its social/cultural significance and its network structure and characteristic has been researched \cite{zlatic2006,palla2008,capocci2006,capocci2008}, including the time evolution of basic properties \cite{buriol2009}.
The policy of the Wikimedia foundation to make the history of all editorial changes visible and to provide an aggregation of the whole Wikipedia has created the empirical basis for all of these studies. In particular, large scale data analysis of link structure as well as text mining and natural language processing profited from having these data sets available. We explore this data set further, by treating Wikipedia as a proxy for knowledge in general and investigate it to learn about the dynamics of knowledge.

Until 2004, Wikipedia Foundation did not have a dedicated system for organizing articles. The structure was contained in the direct links between the articles, without any tools to organize them further. Starting June 2004, the Wikipedia added a feature specifically designed for this purpose. Category pages, which were basically collections of links to various articles or other category pages, were introduced for the first time. Through the use of category pages, it became possible to assign articles to categories, and to link these categories among themselves.

In time, this new feature allowed Wikipedia authors to create an elaborate classification system, albeit the results were quite different then classical classification systems. Wikimedia Foundation itself offered no direct guidance in this procedure --- no keyword list or classification framework. Not only was the vocabulary uncontrolled, but also the mutual relations between category pages were not checked. Thus the ordering of the categories emerged without an externally given template or skeleton.\footnote{An expert in traditional making of reference works such as dictionaries would say that it is impossible to obtain any meaningful ordering this way, if only because of the different meaning of terms for native and non-native speakers as well as for people coming from different cultural contexts. Based on a conversation with Dr. J\"urgen Scharnhorst, a linguist who also worked on projects such as {\em Deutsches W\"orterbuch} of the Grimm brothers and the {\em W\"orterbuch der deutschen Gegenwartssprache}.} The results of this spontaneous, self-organized and collective tagging action at Wikipedia article pages is an interesting grass-root evolution of a category system, unlike any other experts-created system.

While the topical category system of Wikipedia as such has been both analyzed and visualized to some extent, the evolution of the grassroot tagging of categories into a more tangible classification system is a topic that did not attract any attention among the countless studies done about Wikipedia. So far, only a few studies focused on the topic coverage of Wikipedia. Holloway et al. \cite{holloway2007} compared the top categories and the classification structure of Wikipedia 2005 to widely used encyclopedias like Britannica and Encarta. Halavais et al. \cite{halavais2008} evaluated the topical coverage of Wikipedia by randomly choosing articles, manually assigning categories to them, and mapping the distribution of these to the distribution of published books. Wartena \cite{wartena2008} compared clustering obtained through clustering algorithm of the link network with clustering obtained through keywords. A more recent study by Kittur et al. \cite{kittur2009} analyzed the growth of categories, and developed an algorithm to semantically map articles through its category links to the 11 top categories chosen by the research team.

Our work focuses on the evolution of the categories, so it is partly similar to Kittur's analysis, but with some crucial differences. We have focused on the network topology and did not use any semantic analysis in this work. Most importantly our work follows the evolution of the category system as a whole, with a monthly resolution and focus on the dynamics. We also treat the category system differently, without reducing it to a tree.

In a parallel work on the same dataset, we used some simple word matching and attempted to compare and map Wikipedia to other knowledge order systems such as the Universal Decimal Classification (UDC), which is mostly used in libraries \cite{KSL2011a,KSL2011b}.

Our leading question is how the structure of knowledge emerges and changes over time. We treat the Wikipedia as a proxy for knowledge in general, given its extremely broad topical coverage. By investigating the way this knowledge is organized we are trying to find some general trends in the development of knowledge and its perception. As mentioned earlier, the category system has been already investigated \cite{muchnik2007}, but the focus on its evolution is new.

It is possible to analyze the evolution of category organization in many different ways. We wanted to see the structure of knowledge and how it is perceived, so we elected to use the pages and links as our main object of analysis. We treat them as a network that represents knowledge (in case of articles) and organization of this knowledge (in case of categories). We have to emphasize that we did not applied any NLP tools or text analysis, which allowed us to bypass the semantic relevance of category/article page names as well as the content of articles. In essence, we trust in Wikipedia's many authors' insight and that they understand the categories and articles they edit better than any automated system could and that the links they create are relevant.

We do not limit ourselves to strictly looking at the category system, but also want to see how well this system actually works --- whether it describes the actual knowledge well. We have analyzed if the grouping of articles according to the category system reflects well enough the natural grouping of articles into clusters of densely connected (and so closely related) topics. Szyma\'nski \cite{szymanski2010} also performed an analysis of how category network of the Wikipedia and article network match each other, but the study concerned itself with single static snapshot, and the methodology used encounters issues when used to investigate changing system over time.

We explain our approach more detailed in section \ref{sect_data}, where we describe the nature of the data, and how we have processed it to fit our needs. Next, in section \ref{sect_analysis} we detail the analysis methods, and continue with section  \ref{sect_results} where we presents our findings and interpretation of our results. In section \ref{sect_conclusions} we summarize our findings and offer concluding remarks.

\section{Data}
\label{sect_data}

The backups of the whole Wikipedia are created regularly by the Wikimedia Foundation. These database dumps are publicly available.  We have used a dump of the Wikipedia from 2008-01-03.\footnote{2008-01-03 translates to 3rd January 2008. This is the format Wikipedia uses, and hence we will follow the same YYYY-MM-DD format for all dates.} When this research has begun, this was the most recent dumb containing all articles with all revisions ever made. Due to size problems the full Wikipedia dumps were not created successfully after 2008-01. The practice was restored only in 2010.

Since the Wikimedia Foundation keeps dumps in publicly accessible space for a limited time, the older dumps cannot be downloaded from the regular Wikipedia dump link at \verb'dumps.wikimedia.org'. We have therefore obtained the dump file \verb'enwiki-20080103-pages-meta-history.xml.7z' from \verb'http://www.archive.org/details/enwiki-20080103', which is an archive site. Our dump is an xml file that contains all Wikipedia pages and their histories. For each page, there are separate entries for all revisions. Each revision contains the full text of the article at the specific time of when that revision is made. The file, uncompressed is $2.8\mathrm{TB}$ in size. We have split the file into smaller files for processing. Each of these $267$ files have articles with their full revisions and are usually around $10\mathrm{GB}$ in size. We also acquired \verb'enwiki-20080103-page.sql.gz' file from the same source to set up a database for translating page names to page IDs (this process is explained later).

Each Wikipedia page has its own unique ID (an integer number) and name. We wanted to obtain a dynamic network of links between pages and elected to identify them by their IDs. The informations about each page IDs and names are contained in the page entries in the xml file. The information about their links are only found in the text, referenced by their names. This makes is necessary to translate each encountered link from name to ID.

\begin{figure}[ht]
  \epsfig{file=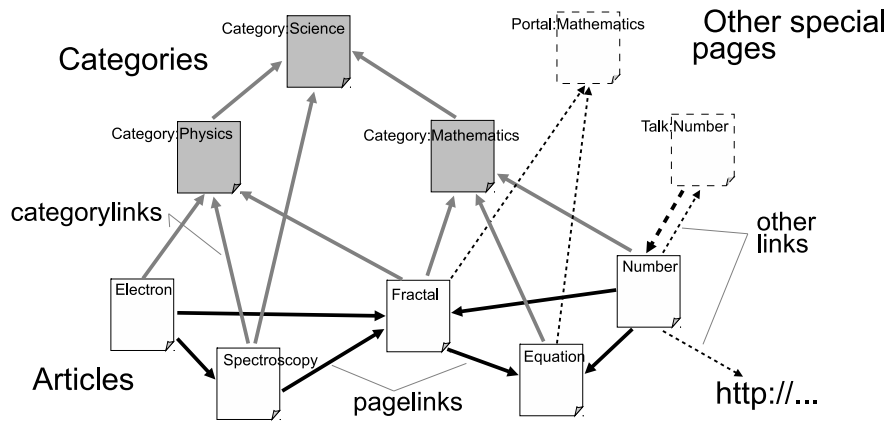,width=\columnwidth}
  \label{explanation1}
  \caption{Illustration of Wikipedia's organization of different pages, showing articles (white), categories (gray) and other pages (white, outline broken) as well as pagelinks (black), categorylinks (gray) and other links (broken line arrows). The page names are only for illustration and the figure does not represent any actual links in Wikipedia. Note that the link from Talk:Number to Number is a pagelink since it targets an article, but it is discarded further on along with the page it originates from. Everything marked in broken lines is discarded.}
\end{figure}

Our first step was to extract the links from the article texts. We distinguished between 3 type of pages, and 3 type of links. The pages are {\em articles}, {\em categories} and {\em other}, while the links can be {\em pagelinks}, {\em categorylinks} and {\em other links}. Since we scrutinize a dynamic system and its history, we also need to acquire the exact time for the links appearances and disappearances. Figure \ref{explanation1} illustrates the relation between all types of pages and links which we have considered.

Articles are normal pages in Wikipedia, that contain information about a subject, such as \verb'Electron', \verb'Electric charge', \verb'Albert Einstein' or \verb'Fun'. Categories are special pages that are used to group articles together. Their names always start with \verb'Category:', such as \verb'Category:Leptons', \verb'Category:German physicists' or \verb'Category:Recreation'.

Other pages include all page types which are not build to contain and share encyclopedic knowledge, but rather to organize it. Such pages include discussion pages attached to articles (\verb'Talk:Electron'), Wikipedia portal pages (\verb'Portal:Physics'), help pages (\verb'Help:Link'). In general all {\em other} pages are those that have one of the recognized special page prefixes (such as \verb'Portal:', \verb'Talk:', \verb'Help:', etc.).

Pagelinks are links from any Wikipedia page type to Wikipedia articles (such as the article \verb'Electron' referring to \verb'Electric charge' article). Links to articles from any page type is technically a pagelink, but since we discarded all special pages other than categories, we are only left with links between articles and links form category pages to article pages. Moreover since we treat categories as containers for articles and other categories, we later ignore all pagelinks from category pages to articles. These are usually links to article pages with the same name as the category page, such as \verb'Category:Physics' linking to \verb'Physics'. The category pages technically can have text with pagelinks, but in practice they do not, as this is not the purpose of those pages.

Categorylinks are links from articles or category pages to category pages (such as "Electron" linking to \verb'Category:Leptons'). It is important to bear in mind that articles or categories can belong to a category through the existence of categorylinks. They are displayed at the bottom of the article pages and are meant not as a normal kind of link, but more in the sense of an assignment of an article to a category. Thus, if the article \verb'Electron' has a categorylink to \verb'Category:Leptons' we treat it as if \verb'Electron' article belongs to the category \verb'Category:Leptons', not just referring to it. This is what we mean by "treating categories as containers of articles". As we have noted before, articles can belong to more than one category, and they often do. More importantly, categories can also belong to another category, and to more than one.

Other links are all other different types of links found in Wikipedia, from links to discussion pages, interwiki links (like links to the same article in a different language), external links and links to pictures. We disregard all these, and thus we work on a set of articles and categories that are linked by pagelinks and categorylinks.

The first step is the extraction of the link data from the article texts. We have written a Java program for this specific purpose, and named it LinkMiner.

All pagelinks and categorylinks are enclosed in double square brackets in the XML text of the dump, such as the link to ``Electric charge'' being \verb'[[Electric charge]]'). The target page of the link is a page with the given name. However, before the name is matched, the link name is formatted according to Wikipedia's own canonization rules. These involves substitution of whitespaces with underscores, removal of leading, trailing and multiple underscores and capitalizing first letter (the rest is not changed, even if capitals). It was necessary to canonize the names of the links ourselves before attempting to translate them into IDs by using the database, otherwise links which are valid in actual Wikipedia would not be recognized. As example, a page could contain link \verb'[[  electric charge]]' introduced by somebody. It would be recognized by Wikipedia engine as referencing \verb'Electric_charge', so the link would exist in Wikipedia. If we did not canonize names before finding correct page just like Wikipedia engine does, we would miss such link.

Wikipedia allows authors to put standardized parts of the articles, like infoboxes, in the pages by only referencing an appropriate template. Such templates may contain links on their own, and these links can actually be found in the Wikipedia articles when browsing. However, in the XML file, these type of links are not part of the page texts. Moreover, if a template contains a link, it is a standardized link that will be found on all other pages that uses the same template, and such links are not necessarily related to the article itself. For this reason, and for simplicity in the processing we decided to omit these links.

All other pages and other links were discarded at this point and do not appear in the further processed data. Now we are left with a list of all revisions of all articles.
For each article, we have the ID and name of the revised pages, timestamp of when the revisions were made, IDs and names of the authors who made the revision and finally the list of links that were added and removed during this revision. The link list is a list of IDs of the target pages, with 0 substituted for unresolved page names. These are the so-called ``red links'', i.e links to non-existing pages, or links to pages that are deleted. The dump does not contain information about deleted pages, but deleted pages are rare occurrence.

We have created two datasets containing extracted revisions, one with only pagelink changes, and second with only categorylinks changes. The split into separate sets for pagelinks and categorylinks is for the convenience, since further on we analyze pagelinks and categorylinks separately and only compare them in the end.
The following example line from pagelink change dataset:\\
{\footnotesize \verb'12 2002-05-01T14:18:43Z 372 Eclecticology ; 0 58198 18280 ; 30925 12988228 93812 ;'}\\
contains a revision of page with ID 12, the revision was made on 1st May 2002, ad 14:18:43 UTC (all timestamps in Wikipedia data is in UTC time), the author was user with ID 372 and name Eclecticology. The revision added 3 links, to pages with IDs 58198, 18280 and one whose ID could not be determined, because it did not exist at all at the time the dump was made. Moreover 3 links were removed, to pages with IDs 30925, 12988228 and 93812. The example is from the set of revisions containing pagelinks information. Since a revision can change both pagelinks and categorylinks, each revision can be found in both sets. The same revision as previous example, but from the set containing categorylinks information is empty --- no categorylinks changes were made:\\
{\footnotesize \verb'12 2002-05-01T14:18:43Z 372 Eclecticology ; ; ;'}\\
The fact that it contains no categorylinks changes is obvious, since the category system was not present in 2002, so no categorylinks were created or removed.

At the end of this procedure, we have two separate datasets: one containing pagelinks changes information and the other containing categorylinks changes information. Both sets together form one dynamic network with two different kinds of vertices and links. This network is continuously changing, with hundreds of edits every second. A significant part of the revisions are simple change-reversal pairs, where a page is changed (quite often vandalized) and reverted shortly after. To be able to trace global structural changes in this network over time we elected to look at the network at a discrete points in time and compare the changes between such snapshots.\footnote{In this paper we do not show the full network structure. For a limited visualization of the structure of the 2008 snapshot see \cite{placesandspaces}.}

We decided to create monthly snapshots, at the first day of each month, exactly at 00:00:00 UTC time. This choice is to a certain extend arbitrary. The first Wikipedia article appeared on 2001-01-16 and first pagelinks appeared shortly after on 2001-01-21.\footnote{It is possible that first article/link were different, but were afterwards deleted completely during early re-organizations. Wikipedia does not keep information of completely deleted pages in the dump --- only dumps from before the page was deleted would contain information about them.} Therefore our first snapshot is from 2001-02-01, while the last one is from 2008-01-01. As the category system is introduced only on 2004-05-30, the categorylink snapshots starts with 2004-06-01.

The snapshots are static graphs with vertices representing both category and article pages. By keeping the list of all article names and IDs we know which vertices represent articles and which represent categories. Wikipedia allows links to non-existing articles. That means that in some cases a link to a page may be created before the actual page. In Wikipedia these are "red links" where a reference to a not-yet described concept is made. Since we are using global name and ID lists, we can translate the destinations for these links, but we discard them when making snapshots. Only links between pages that existed at the time the snapshot is made are taken into account in the analysis.

We create two snapshots for each month --- a pagelinks snapshot, where vertices are articles and links are regular links between articles, and a categorylinks snapshot, where vertices are both articles and categories and links are categorylinks. Note that in categorylinks snapshots the articles are never linked to -- while categorylinks often are present in articles, they always point to categories and therefore no pair of articles can be connected by a categorylink.

To facilitate scientific investigation of the same dataset by others, the data set is publicly available at EASY, an archiving system of Data Archiving and Networked Services \cite{EASY}.

\section{Analysis}
\label{sect_analysis}

Our aim was to analyze the Wikipedia category system as a classification system. Traditional knowledge organization systems are hierarchical structures, with clearly defined division of the whole knowledge within scope into progressively smaller and more specific parts. They can be represented as a hierarchical tree network that have a single \emph{root} vertex that represents all the knowledge with successive layers of vertices representing more and more specific classification categories. Directed links go from a given category to a higher ranking category in the hierarchy.

The Wikipedia is different than a classical knowledge organization system, as it has no specified root and no hierarchy in its category system. Without any point of reference or ``top'' in the system, it is extremely difficult to investigate it as an organization system. At most it could be treated as a directed network and analyzed as such. To see the whole picture more clearly, and not in only statistical terms, we decided to hierarchize Wikipedia category system. This process turns Wikipedia into a structure to knowledge organization system, thus allowing us to analyze it as one. Hierarchization requires pinpointing a root and top level categories in the Wikipedia and building a hierarchy from there.

Unlike classical classification systems, the grassroot organization of Wikipedia category system does not offer top level categories, not in the same sense that the classical knowledge organization system generates. If one looks at the categorylinks, one may define the root category as one that does not belong to any other, but in turn encompasses indirectly all Wikipedia. For the 2008-01-01 snapshot, this root category is the \verb'Category:Contents'. However, its immediate subcategories contain \verb'Category:Wikipedia administration', \verb'Category:Portals' or \verb'Category:Wikipedians' which are not related to actual contents of articles, but to technical aspects of Wikipedia or to methods for knowledge organization. As we are not interested in Wikipedia's administrative and mechanical system, we have decided to search for a more suitable root. This search resulted in the choice of \verb'Category:Main_topic_classifications' as our root category for 2008-01-01. The subcategories of this root category are all topical categories, and it indirectly contains most of the Wikipedia articles.

When we checked the revision history of our root category, we found out that it was introduced only in 2006-10-08. Since we are interested in the time evolution, we needed to find the root category for every snapshot. For the sake of continuity, we decided to consider \verb'Category:Main_topic_classifications' as our root for the whole time it was present in Wikipedia. What became an issue was finding its predecessor. To achieve this, we have looked at its immediate neighbors (ego-network). We have established that the \verb'Category:Fundamental' was the most appropriate topical root category before --- it contained several main topical categories that were moved to \verb'Category:Main_topic_classifications' when it was created, and all its subcategories were broad topical categories as well. It has existed since 2004-06-13, which is 14 days after category pages were introduced. We did not find any predecessor of \verb'Category:Fundamental' --- the category system at that time was rather chaotic and unorganized, since the process of generating categories were a very recent implementation in Wikipedia.

\begin{figure}[ht]
  \epsfig{file=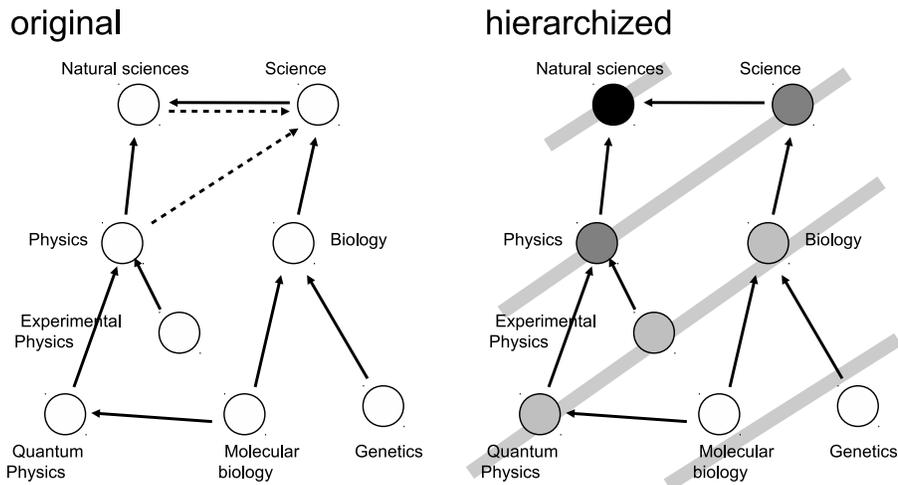,width=\columnwidth}
  \label{explanation2}
  \caption{Illustration of the hierarchization process of an example category network with the root category Natural sciences. On the right hierarchy levels are represented by shade of gray (black for root, white for lowest hierarchy level). The two links drawn with broken lines were discarded because they were not conforming with the established hierarchy. Note that this figure shows only example network and does not represent real Wikipedia links or categories.}
\end{figure}

We imposed the hierarchical order in the network in the following way. Our root category (first \verb'Category:Fundamental', later \verb'Category:Main_topic_classifications') became level $0$, its immediate subcategories (and eventual articles) became level $1$, and so on. Basically, each article's level (or depth) is equal to the shortest directed path length from that article to the root category. Note that all categorylinks are directed from the tagged page to the category. In our scheme, level $2$ categories all have links to level $1$ categories, which in turn have links to the root (level $0$) category. Figure \ref{explanation2} illustrates this hierarchization process.

After assigning levels to all pages, we have discarded all categorylinks that did not conform to the hierarchy. Thus links that connect pages on the same level or links from higher hierarchy levels to lower hierarchy levels have been discarded. In this way, a network without directed loops has been created. This network is still not a tree like classical knowledge organization systems, since a page may belong to more than one category. Since we have a hierarchy in place, we can define \emph{up} and \emph{down} in the hierarchy. Up means from larger depth value to smaller depth value. Categorylinks are always directed this way. Down obviously means opposite, from smaller depth to larger depth. Note that we have processed only the categorylinks network in this way. The pagelinks network was not hierarchized or altered in any way. Since articles can be found at different depths in the categorylinks hierarchy, pagelinks can connect pages found at different depths in this network. Any depth can be connected to any depth with a pagelink.

The first measure of the category structure was the distribution of top category sizes across time. The top categories --- direct subcategories of the root category, are those that are assigned depth $1$. In order to measure the size of these categories, we have used $4$ different measures: number of articles by fractional assignment, number of categories by fractional assignment, number of unambiguously assigned articles and number of unambiguously assigned categories.

\emph{The number of articles by fractional assignment} is the total number of articles that can be found directly or indirectly to belong to a given category. We have used an algorithm where we assign each article weight $1$ and then pass that weight up the hierarchized category network. Categories were used to pass the weight up the hierarchy, but did not contribute to the weight. Since an article may belong to more than one category, we have used fractional assignment. Thus, a single article that belongs to $3$ different categories contributes weight $1/3$ to each of these categories. This weight is then passed on in a similar way upwards: For example, if a category with a weight $3.5$ belongs to $2$ categories, then it will contribute weight $1.75$ to these categories. This process proceeds upwards through the hierarchy, until it reaches the categories at the level $1$.

\emph{Number of categories by fractional assignment} follows the same rules as the number of articles by fractional assignment, except it counts categories (including itself) and not articles. The difference is that all categories are assigned initial weight $1$ that is passed up the hierarchy, while articles are given weight $0$.
When using these methods each article or category will generate only a total of weight $1.0$ weight. Thus the total sum of weights at any specific level is equal to the actual number of articles or categories found in the hierarchized network below that level. When considering top categories --- categories at level $1$, the total is equal to the number of all articles or categories.

\begin{figure}[ht]
  \epsfig{file=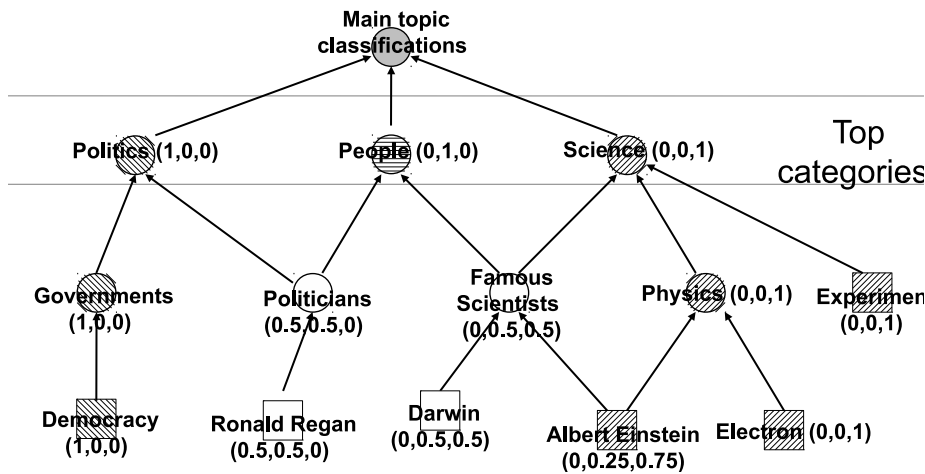,width=\columnwidth}
  \label{explanation3}
  \caption{Illustration of the unambiguous assignment process for an example network. The top categories have weights $1.0$ for themselves and those weights are propagated down the hierarchy to create category weight profiles for each category and article. Each category or article that has more than $0.5$ in one weight (they are normalized to $1$) is assigned to top category that this weight represents. Different category assignments are illustrated as different hatching of the pages, while ambiguous are left blank. Note that this figure is for illustration only and does not represent links or pages in real Wikipedia.}
\end{figure}

\emph{The unambiguous assignment of pages} is a more complex process as it is illustrated in Figure \ref{explanation3}. First, for each snapshot we have a set of $D$ top categories. We define each page (both categories and articles) in the whole network to have a set of $D$ weights. Each top category $i$ ($i \in (1,D)$) is assigned value $1.0$ in $i$-th weight, so each weight $i$ correspond to one top category. Each category then passes down its full weight set to its subcategories and articles that belong to those subcategories. The contributions are added up and normalized so that the sum equals $1.0$. This process is repeated for the next level/depth and the weight sets are thus propagated downwards the hierarchy until they reach all the articles.

When we examine the weight set of any given page (i.e. an article or a category page), we look if any single weight in the set is larger than the threshold value $0.5$. If we follow the example given in Figure \ref{explanation3}, we have 3 top categories, hence the weight sets at the first level are (1,0,0) for \verb'Politics', (0,1,0) for \verb'People' and (0,0,1) for \verb'Science'. In order to decide if an article or a category belongs to which top category, we check the single weight values for that article. Thus, the article about \verb'Albert Einstein' with the weight set of (0, 0.25, 0.75) will belong to the top category \verb'Science'. Since the weight sets are normalized to sum=$1.0$, only a single weight may be larger than $0.5$, and thus any given page can belong only to a single top category. The pages that do not have a dominant top category (no weight exceeds $0.5$) are not assigned to any category and treated as ambiguous. In Figure \ref{explanation3}, \verb'Ronald Regan' and \verb'Darwin' would be ambiguous pages.

We have used 4 different methods to measure the size of a category. In practice, we have found out that the results are similar for all of the methods. We have therefore concluded that all 4 methods describe the same dynamics in a somewhat different way. We have therefore made further conclusions based only on one measure and simply checked if the other measures don't disagree with them. We have chosen the number of articles by unambiguous assignment as our primary measure, since it is used for the modularity.

The second issue we have investigated was the relation between the category system of Wikipedia and the articles that this system describes. On the one hand we have a classification system -- a hierarchized category system, and on the other hand we have a network or direct links between articles. Existence of a direct link shows that there is a topical relation between articles, since completely unrelated articles would not refer to each other. Thus the pagelinks network is a network of relations between articles. We have investigated how well the hierarchized category network groups articles that are related.

We have considered the pagelinks network, where each article is assigned to one of the top categories. We have used unambiguous assignment of articles to decide which article belongs to which category. This means that we have a pagelinks network with a grouping of vertices defined by the structure of the category network. We have looked how such clustering relates to actual linking patterns. We have investigated the modularity value of the pagelinks network in each snapshot, using the clustering defined by category network. We have not used any clustering algorithm, because such algorithms work on each single snapshot separately, while there is a continuity with a \verb'Category:Physics' being the same in January 2008 and in December 2007. Clustering each snapshot separately would cause us to lose the fact that we are investigating one changing system and not several unrelated ones. We have therefore elected to only calculate the modularity value for actual grouping and see if the value points of clustering correspond to the actual pagelink structure or not.

We have used the definition of modularity $Q$ as the difference of actual edge numbers inside each cluster and the number expected from a purely random network (Erd\"os-R\'enyi random graph) with the exactly same number of total edges:
\begin{equation}
 Q=\frac{1}{E} \sum_i \left( E_{ii} - \bar{E_{ii}}\right) \label{def_Q}
\end{equation}
where $E_{ij}$ is a total number of edges between articles in clusters $i$ and $j$, $E_{ii}$ being number of edges inside cluster $i$ and $\bar{E_{ii}}$ is the expected number of edges if the network was purely random, equal to
\begin{equation}
 \bar{E_{ii}} = E \frac{N_i(N_i-1)}{N(N-1)}
\end{equation}
where $N_i$ is the number of vertices in cluster $i$ and $N$ is the total number of vertices. Ambiguous vertices, that do not belong to any clusters, are omitted when calculating $Q$ using equation \ref{def_Q}. They still do contribute to the total number of edges $E$ and vertices $N$. The vertices outside the category system are disregarded completely. Their contributions to the total number of edges $E$ and vertices $N$ are also ignored.

\section{Results and discussion}
\label{sect_results}

\begin{figure}[ht]
  \epsfig{file=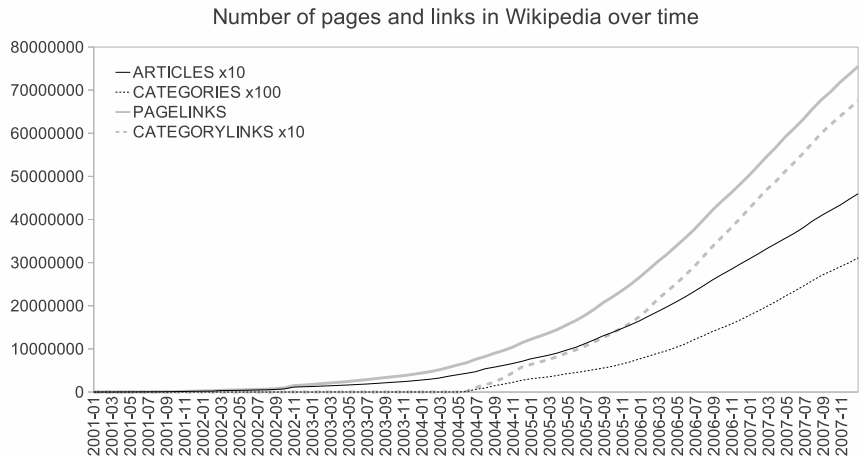, width=0.95\columnwidth}
  \label{grow}
  \caption{The growth of number of articles, categories, pagelinks and categorylinks in Wikipedia over time. It's clear that the number of pages (articles, categories) and corresponding links (pagelinks, categorylinks) is strongly related. Despite starting at different time points, the number of articles and categories also follow the same trend.}
\end{figure}

We have investigated first the general size of the Wikipedia over time, including number of articles, categories, pagelinks and categorylinks. The result is shown in Figure \ref{grow}. The results show us that the number of pagelinks is strongly related to number of articles and number of categorylinks to number of categories. This means that the average number of links per page is not changing much during the evolution.

Our first point of interest is to see the evolution of the top categories in Wikipedia -- the number of top categories and their sizes. As explained in Section \ref{sect_analysis} we have calculated 4 different measures for that, but found them all behaving similar. The Figures \ref{category2} and \ref{categorytable} shows the changes in category sizes over time for number of articles by unambiguous assignment. Other measures behave similarly and were therefore not included in graphs. This is not surprising. Both fractional assignment and unambiguous assignment use the same categorylink network and similar procedure, albeit one going up and one down the category tree. The number of articles and categories follow similar trends (Figure \ref{grow}), so it could be concluded that they are co-evolving and their numbers are closely related. Similarity of measures using article and category numbers reinforce that claim, showing that it is also true on a more local level, not only globally.

Before the introduction of the \verb'Category:Main_topic_classifications', the category system seems to be quite unstable. The reason of this can be explained by the lack of a top-class structure as it is to be found in a classical knowledge organization system. Formally, the top categories were not recognized as such, but rather evolved into an organization under \verb'Category:Fundamental' organically. When the \verb'Category:Main_topic_classifications' was introduced, a more formal classification system was also put in place. After that point in time, the structure of categories can be considered more or less organized. Still, it is possible to observe changes in the top categories, even as late as in 2008. Major reorganizations with sudden and drastic changes in the sizes of some categories, disappearance of others and appearance of new top categories seems to be a part of the dynamic nature of Wikipedia. It should be noted, that while the top categories suddenly ``appear'' on the graph during reorganization, it is safe to assume that these existed beforehand somewhere lower in the hierarchy and only have been moved to the top during the reorganization. This is why one has to be cautious when drawing specific conclusions from these results. It is still possible to clearly see the existence and times of the re-organizations.

\begin{landscape}
	\begin{figure}[t]
    \epsfig{file=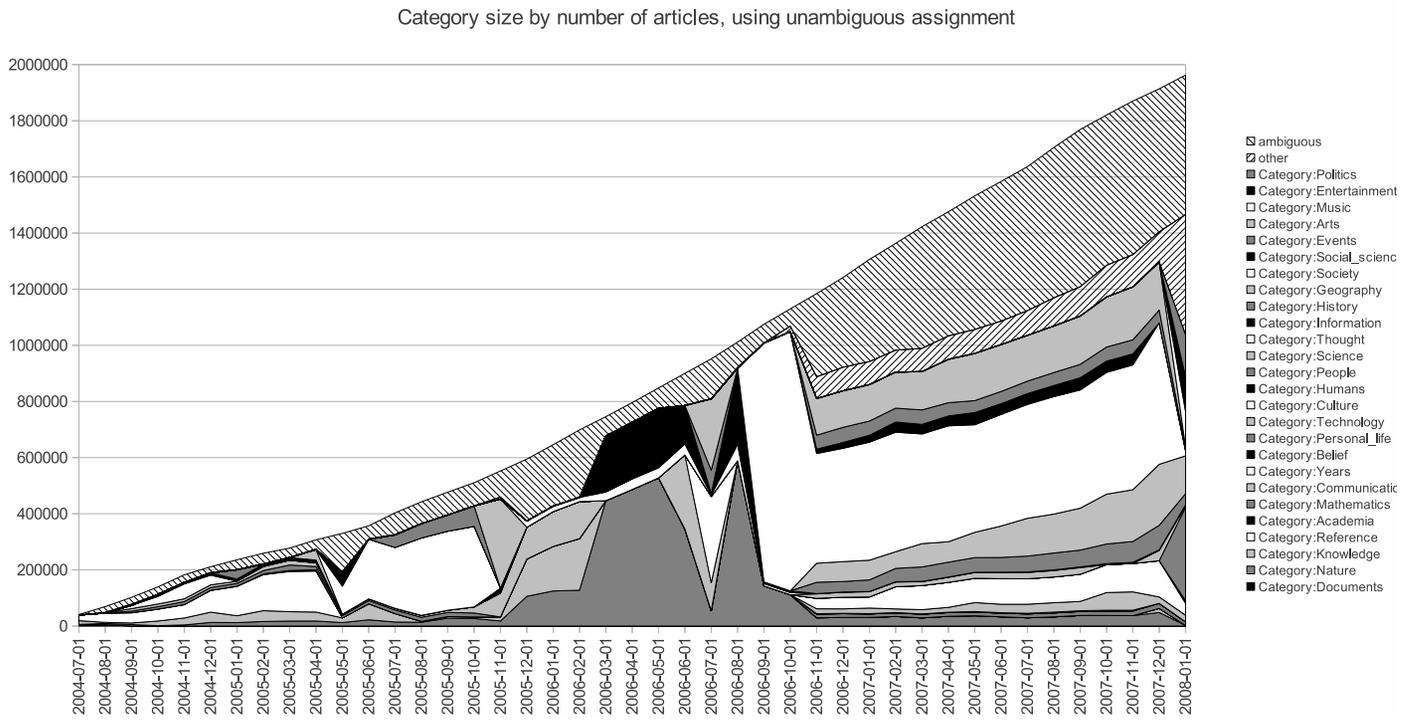, width=\columnwidth}
    \label{category2}
    \caption{The size of Wikipedia top categories according to the number of unambiguously assigned articles. Categories never exceeding $5\%$ share are aggregated into ``other''. The purpose of the figure is to show the dynamics, sacrificing ability to accurately distinguish categories on the graph.}
	\end{figure}
	\begin{figure}[t]
    \epsfig{file=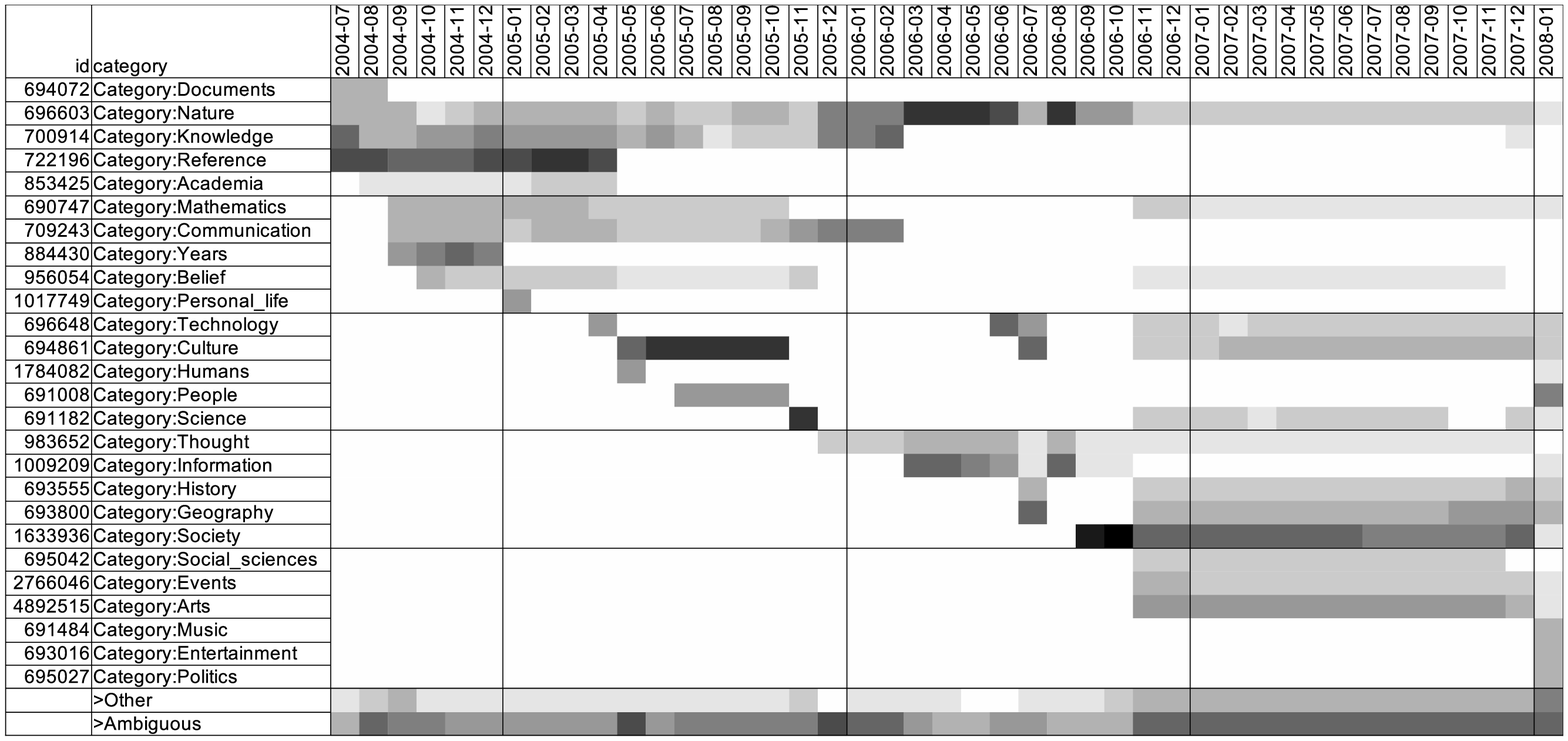, width=\columnwidth}
    \label{categorytable}
    \caption{The relative size of Wikipedia top categories according to the number of unambiguously assigned articles. Shades of grey represent the fraction of af all articles at the given date, that unambiguously belong to the category. The shade scale is square, not linear to differentiate most small values, with thresholds for shades at $0\%$,$1\%$,$4\%$,$9\%$,$16\%$,$25\%$,$36\%$,$49\%$,$64\%$ and $81\%$ (white are categories not present as top categories at given date). The purpose of the figure is to complement Figure \ref{category2} and show which categories were dominant at what time.}
	\end{figure}
\end{landscape}

\begin{figure}[ht]
  \epsfig{file=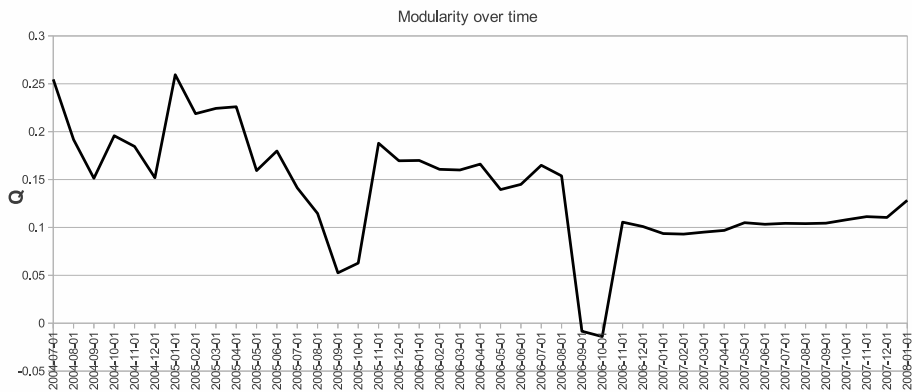, width=\columnwidth}
  \label{modularity}
  \caption{Modularity changes of clustering defined by unambiguous assignment of articles.}
\end{figure}

The relation of the category network to the article network is our second point of interest. For article network snapshots, we have calculated the modularity of the clustering defined by unambiguous assignment (see Section \ref{sect_analysis}). The time dependence of modularity value is shown in Figure \ref{modularity}. The modularity is predominantly positive with values as high as $0.25$. This means that linking under a top category is significantly more frequent than in-between top categories. The large number of vertices assures that the results are not a result of noise. The large changes of the values over time are due to the changes of the category network structure, not the pagelink structure. Since the category network is a hierarchical structure, a small change near the top levels can affect a large number of bottom elements (articles), thus making such shifts possible. Given the rate at which the pagelink network changes, such sudden changes would not have been possible on the pagelinks side.

The modularity declines drastically only in two instances, which happens close to major re-organizations of the category structure. In the first case, it is in 2005-09, before the first major shift, which significantly altered the set of top categories, removing the largest one -- \verb'Category:Years'. The second drop in modularity can be observed shortly before the \verb'Category:Main_topic_classifications' was introduced, in 2006-09. This may suggest that in both cases the category system was in a crisis state, with the mismatch between what categories are supposed to contain and what they actually contained, and that this may have been the cause of the reorganization itself. Another possible explanation is that the reorganizations were already taking place, so that the mismatch resulted from temporary disorder in the category network.

We would also like to point out the following trend: the modularity is on average decreasing, becoming lower after each reorganization. The reason of this can be attributed to one of the top categories being dominant in the early steps, which makes large values easier to attain. Later on, the number of top categories rises, and their distribution become more balanced in size, which makes high values require even higher relative inside-cluster edge densities. This means that the decreasing trend may be a simple artifact of the used modularity measure.

\section{Conclusions}
\label{sect_conclusions}

We have investigated the evolution of the category network of Wikipedia over time. Our research used Wikipedia links, without any form of semantic analysis. Instead we focused on the evolution and changes in the structure. We have treated the category system as a hierarchical classification system with an arbitrarily chosen root category. Our findings show that the Wikipedia category structure is relatively stable for a bottom-up evolved system. We have observed several reorganization events, where the top categories of the system changes rapidly. In between these reorganizations the changes are small.

When the category system is used to group articles into different topical clusters, using a modularity measure we have found that these clusters are more strongly connected internally than in between. This shows that the category system reflects the actual relations between articles, measured by the direct linking between them. We have also discovered that this relation is broken down twice, in 2005-09 and 2006-09, both times preceding a reorganization. After the event, the relation is always restored. Not all reorganizations are preceded by low modularity periods. We have observed that the category system becomes more stable in time, growing more similar to traditional knowledge classification systems, such as Universal Decimal Classification or Library of Congress Classification.

We make analyzed datasets publicly available at EASY \cite{EASY}.

\section*{Acknowledgements}

This work has been performed in Knowledge Space Lab project, a {\it Strategiefonds} project of the Royal Netherlands Academy of Arts and Sciences (KNAW). The data processing has been performed in collaboration with BigGrid-NL, based on a grant of NWO (project ``Emergence of category systems in knowledge spaces - the WIKI case''). We would like to thank in particular Tom Visser, Coen Schrijvers and Ammar Benabadelkader from the BigGrid team for their support. This research has been also supported by the COST action network MP0801 ``Physics of competition and conflicts''.

\end{document}